\documentclass[prl,showpacs,nofootinbib,superscriptaddress,twocolumn]{revtex4}
\usepackage{graphicx}
\def\be{\nopagebreak[3]\begin{equation}}
\def\ee{\end{equation}}
\def\ba{\nopagebreak[3]\begin{eqnarray}}
\def\ea{\end{eqnarray}}

\newcommand{\teta}{\rlap{\lower2ex\hbox{$\,\tilde{}$}}\eta{}}

\begin{document}
\preprint{\vbox{\baselineskip=12pt \rightline{IGPG-06/9-1}
\rightline{gr-qc/yymmnnn} }}
\title{Black hole entropy quantization}
\author{Alejandro Corichi}
\email{corichi@matmor.unam.mx}
\affiliation{Instituto de Matem\'aticas,
Unidad Morelia, Universidad Nacional Aut\'onoma de
M\'exico, UNAM-Campus Morelia, A. Postal 61-3, Morelia, Michoac\'an 58090,
Mexico}
\affiliation{Instituto de Ciencias Nucleares,
 Universidad Nacional Aut\'onoma de M\'exico,\\
A. Postal 70-543, M\'exico D.F. 04510, Mexico}
\affiliation{Institute for Gravitational Physics and Geometry,
Physics Department, Pennsylvania State University, University Park
PA 16802, USA}

\author{Jacobo D\'\i az-Polo}
\affiliation{Departamento de Astronom\'\i
a y Astrof\'\i sica, Universidad de Valencia, Burjassot-46100,
Valencia, Spain}

\author{Enrique Fern\'andez-Borja}
\affiliation{Departamento de F\'\i sica Te\'orica and IFIC, Centro
Mixto Universidad de Valencia-CSIC. Universidad de Valencia,
Burjassot-46100, Valencia, Spain}


\begin{abstract}
Ever since the pioneer  works of Bekenstein and Hawking, black
hole entropy has been known to have a quantum origin. Furthermore,
it has long been argued by Bekenstein that entropy should be
quantized in discrete (equidistant) steps given its identification
with horizon area in (semi-)classical general relativity and the
properties of area as an adiabatic invariant. This lead to the
suggestion that black hole area should also be quantized in
equidistant steps to account for the discrete black hole entropy.
Here we shall show that loop quantum gravity, in which area is
{\it not} quantized in equidistant steps can nevertheless be
consistent with Bekenstein's equidistant entropy proposal in a
subtle way. For that we perform a detailed analysis of the number
of microstates compatible with a given area and show consistency
with the Bekenstein framework  when an oscillatory behavior in the
entropy-area relation is properly interpreted.
\end{abstract}

 \maketitle


Black hole entropy is one of the most intriguing constructs of
modern theoretical physics. On the one hand, it has a
correspondence with the black hole horizon area through the laws
of (classical) black hole mechanics. On the other hand it is
assumed to have a quantum statistical origin given that the proper
identification between entropy and area $S=A/4 \,\ell^2_p$ came
only after an analysis of {\it quantum} fields on a fixed
background \cite{BH}.

Furthermore, it has long been argued by Bekenstein that the
proportionality between entropy and area, for large, classical
black holes, can be justified from the adiabatic invariance
properties of horizon area when subject to different scenarios
(see \cite{cristod} and  \cite{beken2} for a review). Further
heuristic quantization arguments lead to the suggestion that area,
when quantized, should have a discrete,   equidistant spectrum
{\it in the large horizon limit} \cite{beken2},
\be A= \alpha\,\ell^2_p\,n\, , \ee
with $\alpha$ a parameter and $n$ integer. The relation between
area and entropy that one expects to encounter in the large
horizon radius is then extrapolated to the full spectrum. This
would imply that entropy too would have a discrete spectrum, a
property that might also be expected if entropy is to be
associated with (the logarithm of) the number of microstates
compatible with a given macrostate. When this condition is
imposed, then the area is expected to have an spectrum of the
form,
\be A= 4\,\ell^2_p\,\ln(k)\,n\, , \label{areasp} \ee
with $k$ and $n$ integers \cite{bek-muk}. Even when appealing and
physically well motivated, these arguments remain somewhat
heuristic and have no detailed microscopic quantum gravity
formalism to support them.

A quantum canonical description of black holes that has offered a
detailed description of the quantum horizon degrees of freedom is
given by loop quantum gravity (LQG) \cite{ABCK}. This formalism
allows to include several matter couplings (including non-minimal
couplings) and black holes far from extremality, in four
dimensions. There is no restriction in the values  of the matter
charges. The approach uses as starting point isolated horizon (IH)
boundary conditions at the classical level, where the interior of
the black hole is excluded from the region under consideration.
The quantum degrees of freedom are excited when a spin network --a
collection of edges with `spin' labels $j_i$ and vertices--
pierces the horizon, creating punctures that acquire, apart from
the spin $j_i$ endowing it with area, new quantum numbers $m_i$
(responsible for its intrinsic geometry and such that $-|j_i|\leq
m_i\leq|j_i|$). The numbers $(j_i,m_i)$ can be thought of as the
analogues of the total angular momentum and projection along an
axis respectively. These horizon degrees of freedom fluctuate
independently of the bulk degrees of freedom and give raise then
to the entropy of the horizon. There is also an important issue
regarding this formalism. LQG possesses a one parameter family of
inequivalent representations of the classical theory labelled by a
real number $\gamma$, the so called Barbero-Immirzi (BI) parameter
that is absent classically. The strategy is to chose the value of
$\gamma$ in such a way that, for large black holes, the entropy
corresponds to (1/4 of) the area in Planck units \cite{ABCK}.

There is however an obvious inconsistency between loop quantum
gravity and Bekenstein's considerations: the area spectrum in LQG
is {\it not} evenly spaced. On the contrary, the LQG area spectrum
is given by,
\be A=\sum_i 8\pi\gamma\,
\ell^2_p\,\sqrt{{j_i}\left({j_i}+1\right)}\, ,
 \label{area-sp} \ee
where $j_i$ are semi-integers and the sum is taken over all the
punctures $i$ at the horizon. The spectrum (\ref{area-sp}) is not only not
equidistant, but it can be expected that the eigenvalues accumulate
for values of $A$ large in Planck units, given they do for the general
area spectrum \cite{QTGI}.

The inconsistency between loop quantum gravity and Bekenstein's
heuristic arguments seemed to become less relevant when Dreyer
noted \cite{dreyer} that LQG might also be consistent with the
constraints imposed by asymptotically damped quasi-normal modes,
as Hod had previously conjectured \cite{hod} within Bekenstein's
formalism. The idea is that the asymptotic frequency of these
classical modes would correspond to the energy of horizon quanta
through the standard relation $E=\hbar\omega$. This requirement
would then imply that, in the Bekenstein approach, black holes
have an equidistant spectrum  given by
$$A=4\,\ell^2_p\,\ln{(3)}\,n\, ,$$
whereas, in the LQG approach, a {\it minimum area gap}, associated
to the quantum transition, would be given by
$a_0=4\,\ell^2_p\,\ln{(3)}$ (This requirement implies a particular
choice of the Barbero-Immirzi parameter $\gamma$ involving
$\ln{(3)}$). Even when not fully consistent (area spectra
continues to be different), the appearance of a $\ln{(3)}$ factor
seemed to be more than just a coincidence. This initial
expectation was however diminished when it was shown that the
entropy calculation in LQG gave a different proportionality factor
between entropy and area that called for a different value of the
Immirzi parameter that was no longer compatible with Hod's
considerations \cite{Dom:Lew} (See also \cite{GM}).

The purpose of this letter is to show that there is indeed a deep
relation  between entropy within the LQG formalism and
Bekenstein's heuristic picture (supplemented by Hod's
conjectures),  even when the relation is much more subtle than it
was originally conceived. To be precise, we shall show that a
detailed analysis of the number of states compatible with the
macroscopic conditions imposed on small, Planck size black holes
within the LQG approach yields, when appropriately interpreted, a
functional form of the entropy as function of horizon area that
realizes in a precise manner Bekenstein's picture. The coincidence
turns out to be not only qualitative, but it also incorporates two
numbers that are important for both formalisms, namely $\ln{(3)}$
and the value $\gamma_0$ of the Immirzi parameter (that recovers
the Bekenstein-Hawking relation $S=A/4$ for {\it large} black
holes).

\begin{figure}
  \includegraphics[angle=270,scale=.32]{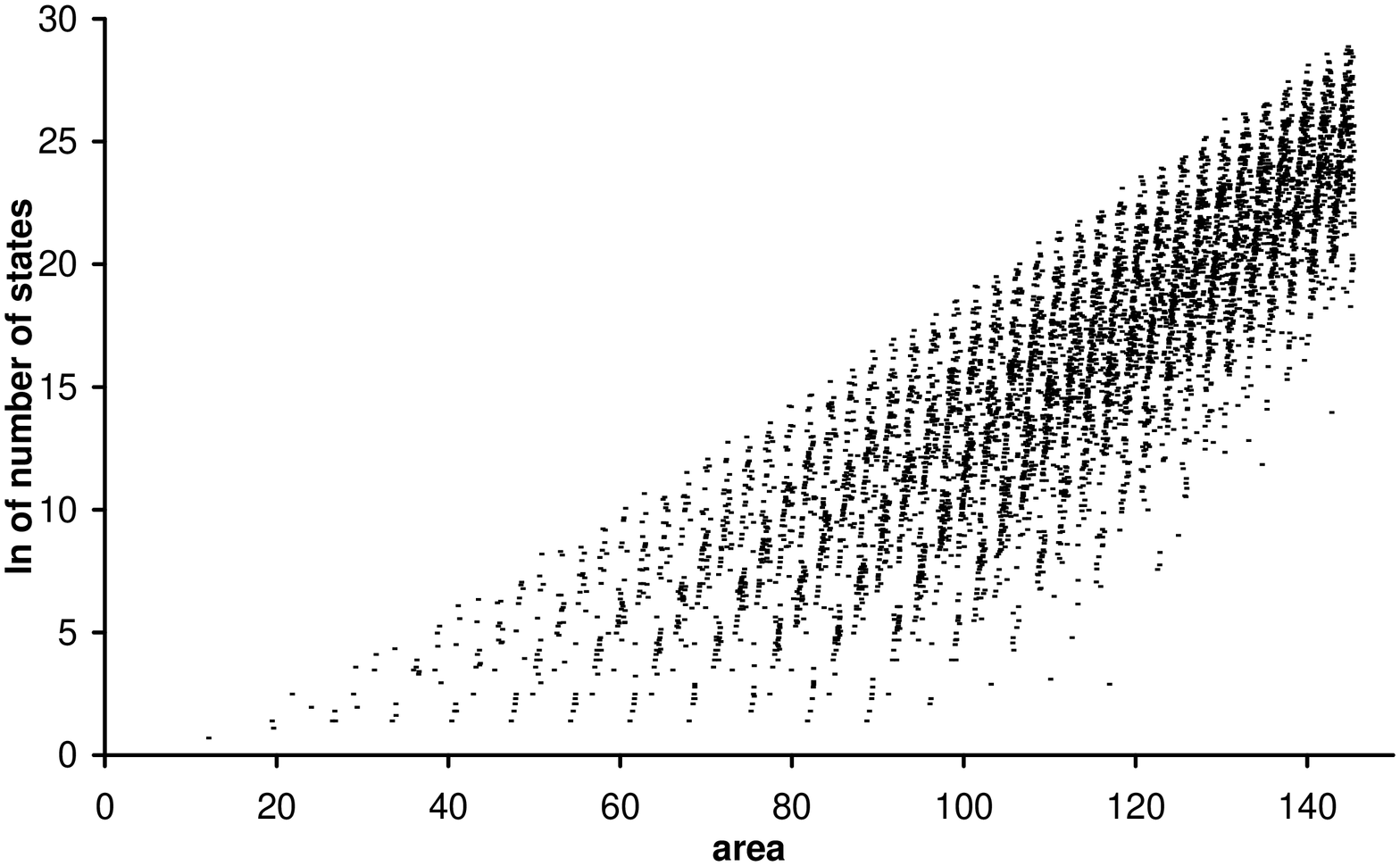}
\caption{\label{fig:1} The ($\ln$ of the) number of states as a
function of area is shown. The Barbero-Immirzi parameter is taken as
$\gamma=0.274$ from \cite{GM} and \cite{CDF}.
The interval $[A_0-\widetilde{\delta A},A_0+\tilde{\delta A}]$ is
taken to be rather small ($\widetilde{\delta A}=0.005\,\ell^2_p$) so that one is
effectively counting the number of states as function of area.
The area is shown in Planck units. Note that this does {\it not} represent the
entropy, given that the interval is very small in Planck units.}
\end{figure}

\begin{figure}
  \includegraphics[angle=270,scale=.32]{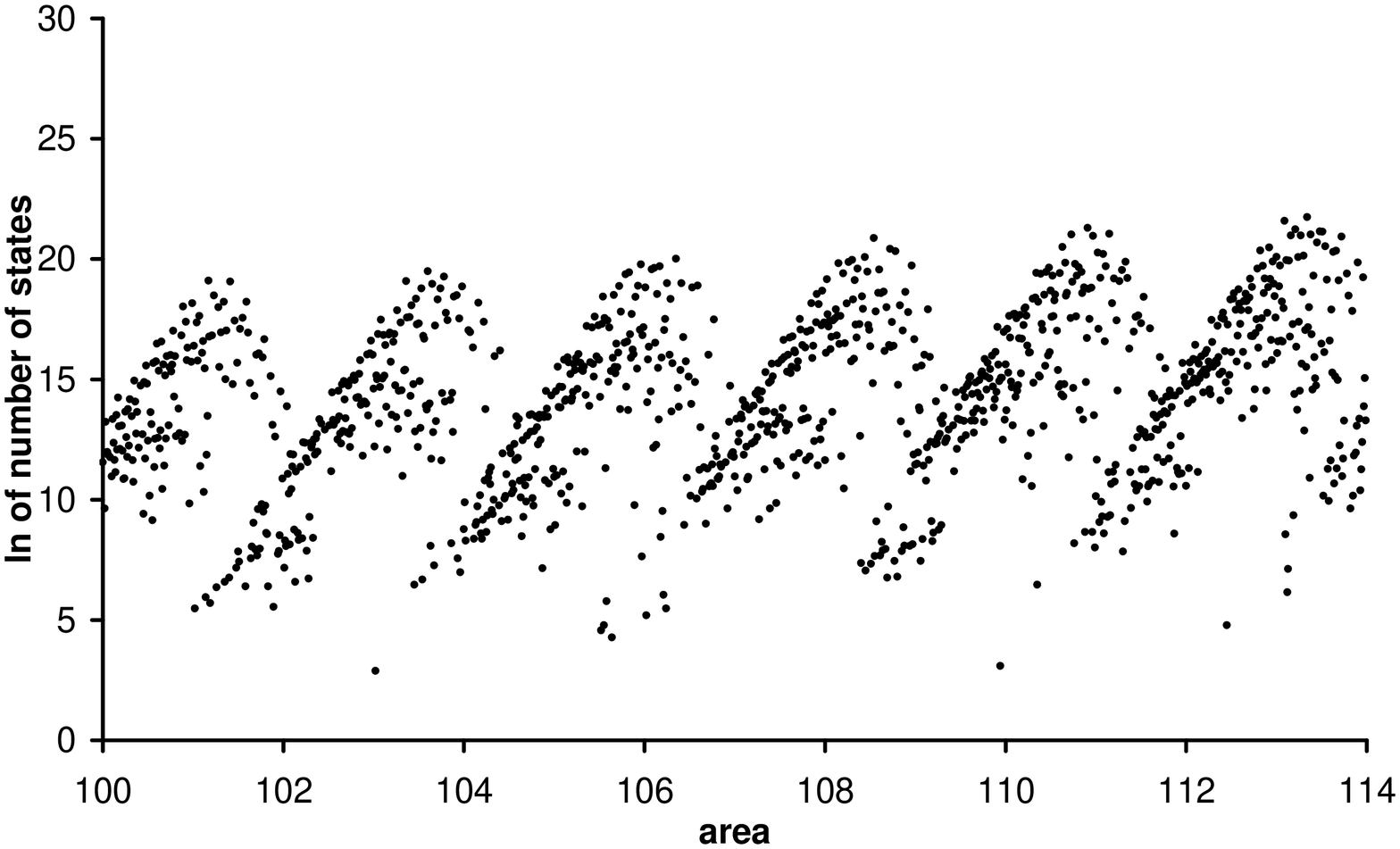}
\caption{\label{fig:2} The same as Fig.~\ref{fig:1} but more detail is shown.}
\end{figure}

We have computed the number of states compatible with a horizon of
area $A_0$ using the formalism developed in \cite{ABCK}, that
specifies which states have to be counted. We performed the
counting using a simple algorithm described in detail in
\cite{CDF}. In the entropy computation within the micro-canonical
ensemble, one resorts to the usual prescription of counting states
whose area eigenvalues $A=\langle \hat{A}\rangle$ lie in an
interval $[A_0-\delta A,A_0+\delta A]$, and where a
total projection constraint $\sum_i m_i=0$ is imposed such that
the horizon geometry is the quantum version of an isolated horizon \cite{ABCK}.
The parameter $\delta A$
that fixes the interval is normally assumed to be of the order of
Planck area. In \cite{CDF} it was shown that the entropy, as
function of area $A$ has some oscillatory behavior, whose
amplitude depends on $\delta A$ but with a constant periodicity that is
independent of $\delta A$.
Here we have taken further the analysis of \cite{CDF} in order to
unravel the structure of these oscillations. As a first step we
have taken a rather small interval $\widetilde{\delta A}= 0.005$ (in Planck units)
with a point separation of $0.01$, in order to isolate the
`spectrum' of the quantum black hole. Note that with this choice,
one is covering the full set of values of area, without the
intervals overlapping, and what one is doing is to separate the
total number of black hole states in different ranges of area, as
is done when drawing a histogram. The resulting number is not then employed
to determine the entropy (for which a much langer $\delta A$ is employed).
The results are plotted in
Fig.~\ref{fig:1} and Fig.~\ref{fig:2}. The oscillatory behavior
found in the entropy \cite{CDF} as well as the patterns shown in
these figures have a period of $\Delta A_0 = 2.41\, \ell^2_p$
approximately.

\begin{figure}
  \includegraphics[angle=270,scale=.32]{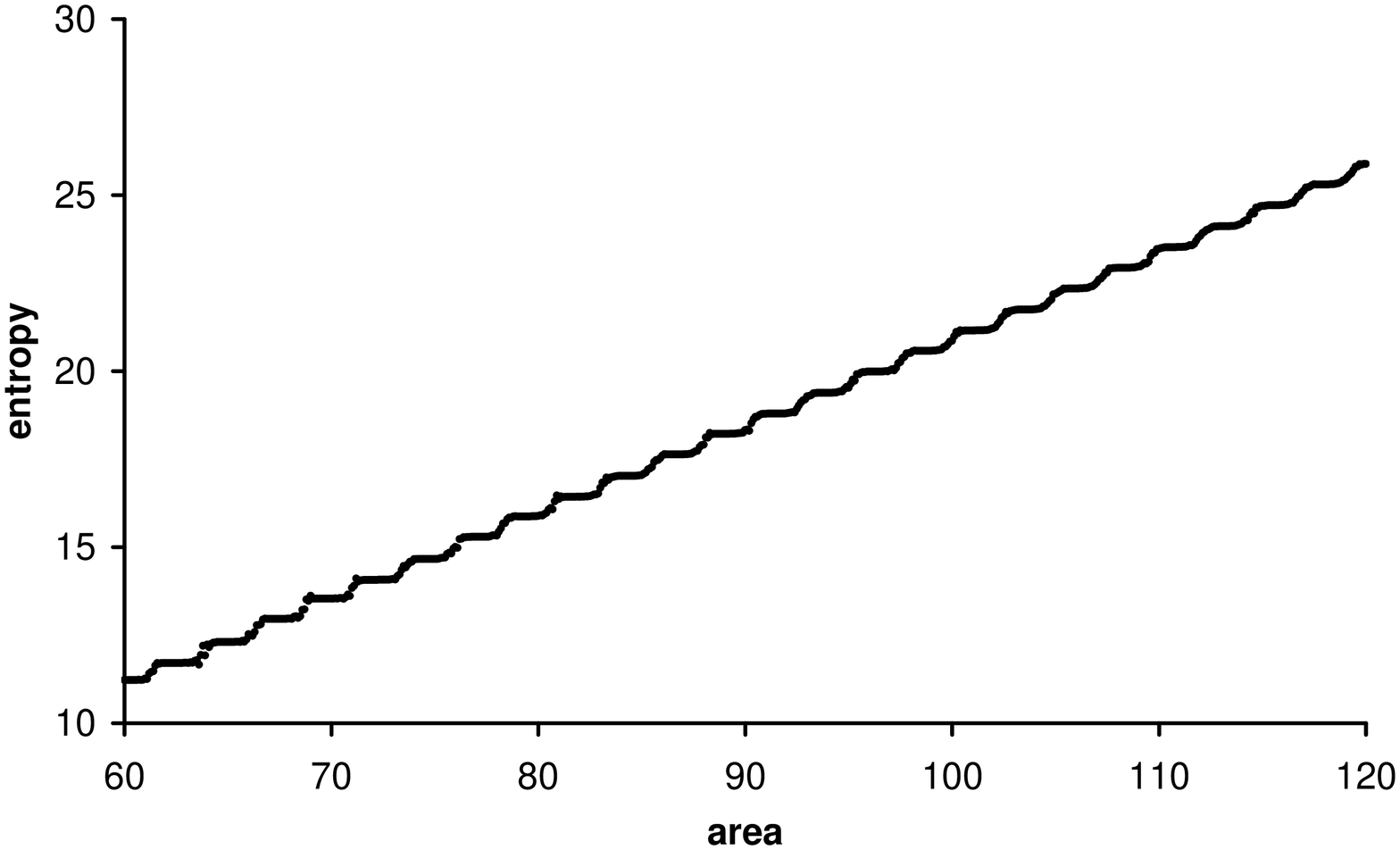}
\caption{\label{fig:3} The entropy as a function of area is shown,
where the projection constraint has  been imposed, the
Barbero-Immirzi parameter is taken as $\gamma=0.274$ and $\delta
A$ is taken to coincide with 1/2 of the period of the oscillations
in the number of states $\Delta A_0$.}
\end{figure}

\begin{figure}
   \includegraphics[angle=270,scale=.32]{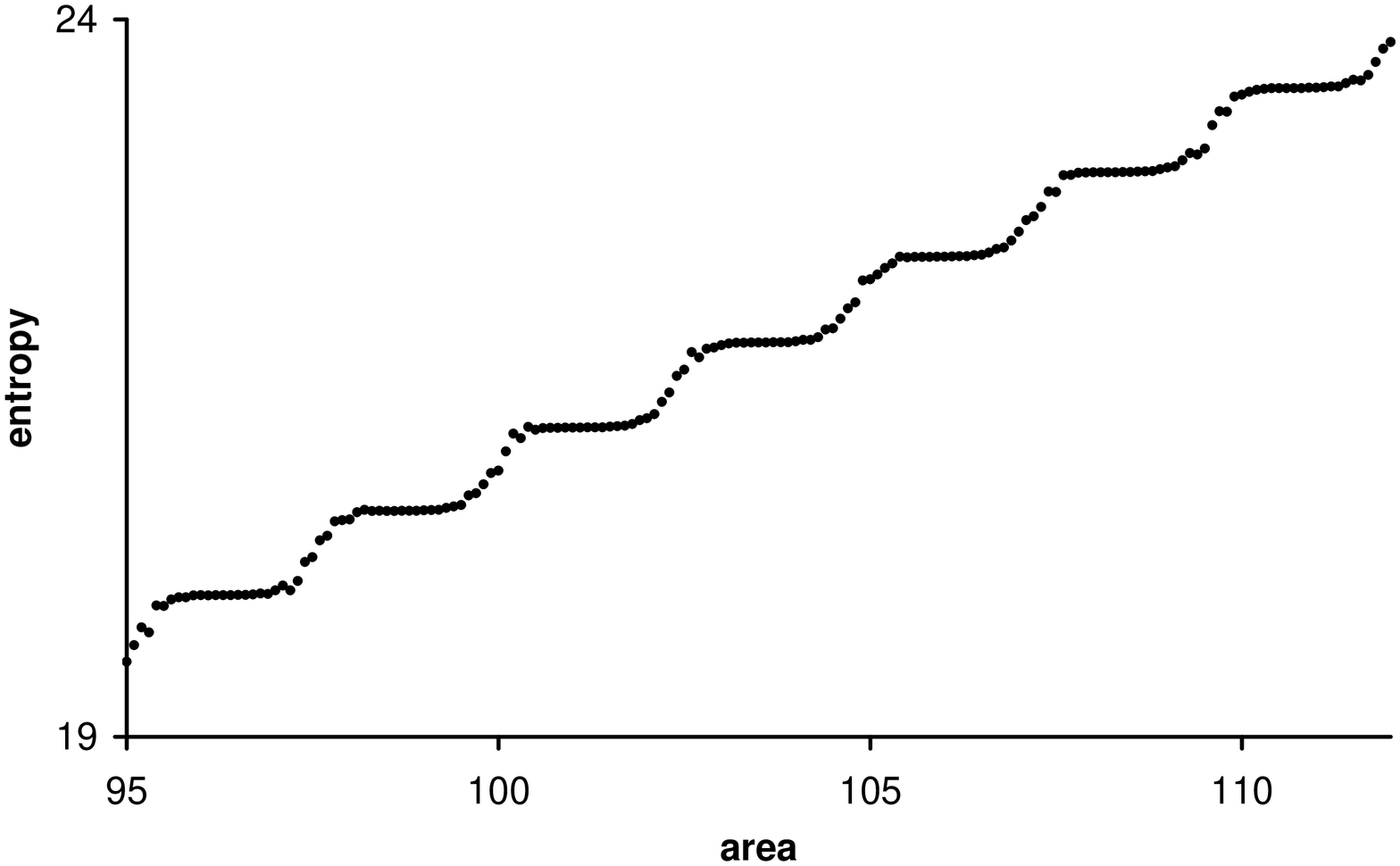}
\caption{\label{fig:4} Same as Fig.~\ref{fig:3} but more detail is shown}
\end{figure}

The next step was to compute the entropy by counting the number of
states within a given interval of area, with the choice that the
size of the interval  coincide with the periodicity of the
oscillations, namely $2\,\delta A= \Delta A_0$. The resulting
entropy is plotted in Figures \ref{fig:3} and \ref{fig:4} where
more details can be appreciated.

Let us now discuss the results. From Figures~\ref{fig:1} and
\ref{fig:2} it is clear that the spectrum of the quantum black
hole has some new and non-trivial features. Specially noteworthy
is the periodic structure that arises when looking at this rather
small scale (recall that each Planck area is covered by 100
intervals and thus correspond to  100 points in the graph). The
appearance of these `mountain like' structures, that are also
periodic with the same period as the oscillations could not have
been inferred from the oscillations in the entropy function. Thus,
the periodicity of the entropy area relation has to be associated
with these new structures in the spectrum and not with other
features such as the change in the number of punctures, a simple
transition involving creation/annihilation of edges puncturing the
horizon, or any other `naive' explanation of that sort. It is
certainly intriguing that this new length scale appears, that as
we would like to emphasize, is not related to any other scale
previously found in LQG.

Motivated by these considerations, it was natural to explore the
entropy counting with an area interval $\delta A$ given by this
new scale $2\delta A= \Delta A_0$. The results, shown in Figures
\ref{fig:3} and \ref{fig:4} are quite unexpected. The oscillations
that are found for all other values of $\delta A$ disappear and
instead, one is left with a `ladder' in the entropy {\it vs} area
graph.

The first conclusion from this graph, is that if one interprets
the (ln of the) number of states as physical  entropy then there
are regions where the area changes but the entropy  remains
constant. Any quantum transition between those states would then
correspond, in a precise sense, to an adiabatic process.
Furthermore, we see that entropy (and not area) has effectively
only a discrete number of possible values it can take. This is
precisely the conclusion that one can draw from Bekenstein's
argument, namely that entropy should be equidistant for large
black holes. Even when it can not be fully appreciated from the
figures, what we observe is that the ladder is not completely
regular for small black holes; the height of the ladder seems to
increase, as the black holes grows, approaching a constant value
for larger black holes. Thus, what we see is an emergent picture
for small black holes within LQG that is consistent with
Bekenstein's model. Furthermore, the manner in which the discrete
equidistant values emerge is much more subtle than just assuming
an equidistant area spectrum. From our perspective, it is a rather
non-trivial result that loop quantum gravity does accommodate
Bekenstein's picture for quantum black holes in such a subtle way.
This is the main result of this letter.

In order to study the dependence of the period of both area and
entropy on the value of the Barbero-Immirzi parameter $\gamma$, we
performed a series of runs of the code with different values of
the parameter $\gamma$. For the area, we found that the period is
indeed linearly dependent with $\gamma$ as has the following
(conjectured) dependence:
\be \Delta A \approx 8\,\gamma\,l^2_p\,\ln{(3)}\, .\label{deltaA}
\ee
The plot in Figures~\ref{fig:3} and \ref{fig:4} were drawn for the
value $\gamma_0\approx 0.274 \ldots$ of the parameter that reproduces the
Bekenstein-Hawking relation $S=A/4$ in the large area limit (see
\cite{CDF,GM} for details). The fact that
the periodicity in area depends on the value of the
Barbero-Immirzi parameter is not surprising since the operator and
therefore its eigenvalues depend on it.

For the entropy, we have made the same estimations and the result
is somewhat intriguing: the asymptotic size of the steps found in
the entropy do not seem to depend on the value of the
Barbero-Immirzi parameter $\gamma$. That is, if the conjectured
numerical value of the area scale (\ref{deltaA}) is true, then
what we find is an universal value in which the entropy is
quantized, namely
\be
\Delta S \approx 2\,\gamma_0\,\ln{(3)}\, .\label{2ndres}
\ee
It is certainly remarkable that, as black holes become large,
entropy seems to be quantized in integer units of a quantity that
contains both `key' numbers: for the heuristic Bekenstein model,
$\ln{(3)}$, and for loop quantum gravity, the value  $\gamma_0$ of
the BI parameter. The precise form of the entropy spectrum is
slightly different from Eq.~(\ref{areasp}) (where $\Delta S=
\ln{(3)}$), but one should also be aware that the relation
(\ref{areasp}) was arrived at by means of plausibility arguments
rather than a hard core derivation. The conjectured entropy
quantization condition derived from (\ref{2ndres}) is the second
result of this letter.

The main features we have found here about the quantum horizon
system, namely the existence of a pattern in the black hole
spectrum with a periodicity that permeates to the entropy area
relation, and the appearance of a new scale associated with this
period, could in principle be `generic'. That is, one might
imagine that these features are common to many quantum systems
with a finite number of degrees of freedom. In order to rule out
this possibility we have repeated the analysis for a quantum
horizon in which the area spectrum is equidistant and given by
${A}'=8\pi\gamma\, \ell^2_p\sum_i j_i$ (an operator that has been
suggested within LQG as well). This would also correspond to the
case (modulo a constant) of $N$ decoupled harmonic oscillators in
the micro-canonical ensemble. Perhaps not unexpectedly, we have
seen that the black hole spectrum is in this case equidistant with
an area separation of $\Delta {A}'=8\pi\,\gamma \,\ell^2_p$, which
corresponds to the increase in area when one adds a couple of
punctures (the projection constraint $\sum_i m_i=0$ prevents one
from having an odd number of punctures that have the minimum
allowed spin, namely $1/2$). There is no non-trivial periodic
patterns in the spectrum and the entropy has discrete jumps that
are directly associated to the fact that the area spectrum is
equidistant.

Another possibility is that this behavior is a consequence of the
particular counting procedure used, and that a different one
\cite{Dom:Lew} might not have the same properties. We have
performed the counting using that procedure and have found the
results to be robust: the entropy has discrete jumps and the
relations (\ref{deltaA}) and (\ref{2ndres}) continue to be valid.
Details will be published elsewhere.

We conclude then that the non-triviality of the loop quantum
gravity area spectrum (\ref{area-sp}) is what brings the new and
unexpected features to the entropy  {\it vs} area relation that we
have reported in this letter, and is therefore responsible for
black hole entropy quantization. Needless to say, these results
can only be a hint of a deeper structure involving gravity,
thermodynamics and the quantum that remains to be unravelled.

\section*{Acknowledgments}

\noindent We thank A.~Ashtekar, J.~Lewandowski and
J.~Navarro-Salas for discussions. This work was in part supported
by CONACyT U47857-F, ESP2005-07714-C03-01 and FIS2005-02761 (MEC)
grants, by NSF PHY04-56913,  the Eberly Research Funds of Penn
State and by the AMC-FUMEC exchange program. J.D. thanks MEC for a FPU
fellowship.

\end{document}